# Process Mining for self-regulated learning assessment in eLearning

*Rebeca Cerezo*, Alejandro Bogarín** and Cristóbal Romero***

**Universidad de Oviedo and **Universidad de Córdoba*

Content assessment has broadly improved in e-learning scenarios in recent decades. However, the eLearning process can give rise to a spatial and temporal gap that poses interesting challenges for assessment of not only content, but also students' acquisition of core skills such as self-regulated learning. Our objective was to discover students' self-regulated learning processes during an eLearning course by using Process Mining Techniques. We applied a new algorithm in the educational domain called *Inductive Miner* over the interaction traces from 101 university students in a course given over one semester on the Moodle 2.0 platform. Data was extracted from the platform's event logs with 21629 traces in order to discover students' self-regulation models that contribute to improving the instructional process. The *Inductive Miner* algorithm discovered optimal models in terms of fitness for both Pass and Fail students in this dataset, as well as models at a certain level of granularity that can be interpreted in educational terms, which are the most important achievement in model discovery. We can conclude that although students who passed did not follow the instructors' suggestions exactly, they did follow the logic of a successful self-regulated learning process as opposed to their failing classmates. The Process Mining models also allow us to examine which specific actions the students performed, and it was particularly interesting to see a high presence of actions related to forum-supported collaborative learning in the Pass group and an absence of those in the Fail group.

At the beginning of the third millennium a new form of learning called e-learning came to stay. We defined it as instruction delivered on a digital device intended to support learning (Clark & Mayer, 2016). However, e-learning can fail in education when overestimating what it can accomplish by itself (Aljawarneh, Muhsin, Nsour, Alkhateeb, & AlMaghayreh, 2010). The benefits gained from this new technology depend on the extent to which it is used in ways which are compatible with human cognitive skills, guided by educational science research principles.

Most of the literature about distance learning has focused on students' achievement outcomes as noted in some of the most important meta-analyses in the field (e.g. Bernard et al., 2004; Cook, Levinson, Garside,Dupras, Erwin, & Montori, 2008; Hattie, 2008; Means, Toyama, Murphy, Bakia, & Jones, 2009; Merchant, Goetz, Cifuentes, Keeney-Kennicutt, & Davis, 2014; Schmid et al., 2009). However, less is known about skill assessment in e-learning; different facets of competences apart from theoretical and methodical knowledge such as skills required for problem solving and personal/social competences (e.g., self-regulation, metacognition, media competence, etc.) (Paechter, Maier, & Macher, 2010). In this regard, those involved in the e-Teaching-e-Learning process do not share a physical interaction space, which can give rise to a gap that raises interesting challenges for assessing students' skills (Lara, Lizcano, Martínez, Pazos, & Riera, 2014). One of those skills is the self-regulation of learning, a process of thoughts, feelings, and actions generated by students, systematically oriented towards the achievement of their goals (Zimmerman, 2013).

Self-Regulated Learning (SRL) processes are particularly important in web-based courses because students are often asked to complete learning tasks with little or no support, requiring them to be highly self-regulated. e-Learning has brought new opportunities to education (European Commission, 2014) but also bring many challenges for the student, who has to to decide what, when, how, and for how long to learn (Sánchez-Santillán, Paule-Ruíz, Cerezo, Álvarez-García, 2016). E-learners have choices regarding the time, place, and the regulation of learning processes, but students at every educational level have deficits in this sense even when they reach higher educational levels such as university (Bjork, Dunlosky & Kornell, 2013). Similarly, there is abundant empirical evidence suggesting that learners do not successfully adapt their behavior to the self-regulatory demands of e-Learning environments (Azevedo & Aleven, 2013; Azevedo & Feyzi-Behnagh, 2011; Cerezo, Esteban, Sánchez-Santillán, & Núñez, 2017; Cerezo, Sánchez-Santillán, Paule-Ruiz, & Núñez, 2016). In short, SRL becomes an essential skill in this context but a challenge in terms of assessment (Azevedo et al., 2013; You, 2015, 2016).

Nowadays, with the development of e-learning, information systems enable us to capture many student actions and interactions from low level events such as mouse gestures and clicks, to higher-level events such as students' learning patterns and processes. These systems have tracking and logging capabilities to gather different types of data such as click streams, chat logs, motion tracking, learning resource usage logs, interaction logs, etc. (Bogarín, Cerezo, & Romero, 2018). Educational Data Mining (EDM) has been applied extensively to these logs in order to discover, monitor and improve educational processes. However, EDM techniques are not generally aimed at discovering, analyzing or visualizing the complete skill process because they do not focus on the process but on the result. To allow analysis in which the process plays the central role there is an emerging line of data-mining research called Educational Process Mining (EPM) (Romero & Ventura, 2013).

The goal of EPM is to extract knowledge from event logs recorded by an educational system, generally through widely used Learning Management Systems (LMSs). These systems are ubiquitous in higher education, with 99% of US colleges and universities reporting that they have an LMS in place (Dahlstrom, Brooks, & Bichsel, 2014). Most of the EPM work has concentrated on supporting company processes in business contexts (Van Der Aalst, 2011) but there is also an increasing body of research in EPM (Bogarín et al., 2018) and a few attempts about SRL in LMSs: Emond and Buffett (2015) applied process discovery mining and sequence classification mining techniques to model and support SRL in heterogeneous learning environments; Reiman et al. in 2014 proposed the use of PM with learning traces based on theoretical principles of SRL; and Bannert, Reimann, & Sonnenberg (2014) detected differences in frequencies of SRL events using PM techniques.

Although there is a large body of previous research in applying EPM, the algorithms reporting quality metrics to address educational issues in the literature are limited to Alpha Miner, Heuristic Miner and Evolutionary Tree Miner (Bogarín et al., 2018). In the present work, we propose the use of a new algorithm for assessing SRL in e-Learning, Inductive Mining (IM), which tends to improve models previously obtained by EPM with other discovery algorithms. Previously used PMs do not return good quality metrics with real event logs where the data is often noisy (Romero, Ventura, & García, 2008). IM is being extensively used in business with very promising results (Leemans, Fahland, & van der Aalst, 2013) and is able to cope with infrequent behavior and large event logs, while ensuring soundness (Leemans, Fahland, & van der Aalst, 2014). Based on that, we want to test its performance in modelling SRL processes.

In short, our aim is to assess students' SRL skill during an e-Learning course through a new EPM technique. SRL assessment is a key but very challenging skill to assess, in both face-to-face and computer based learning environments. We want to check whether PM techniques can contribute to meeting this challenge in e-Learning environments.

Below, we address the study method including the data preprocessing process. Following that we present the results and finally discuss their educational value.

## METHOD

### *Sample*

We used data from 101 undergraduate students (mean age=20.23; SD=1.01; 83% women) from a university in the North of Spain, who completed an online course using the corporate LMS Moodle 2.0 which after preprocessing led to 21,629 events.

### *Pre-processing*

It is necessary to preprocess the Moodle log when we use real event logs (Romero et al., 2008). In general, the log file provided by Moodle contains all of each student's events recorded during their interactions with the LMS summarized in six attributes: The name of the course, the IP of the device used to access Moodle, the date and time they accessed Moodle, the name of the student, the action that student performed, and finally further information about the action.

In this study we used only four attributes: Time, Identifiers(ID) of the students (we converted the students' names into IDs to maintain their anonymity and ensure the principles of ethical and professional conduct), Action and Information. We deleted duplicate records, and instructor, system administrator and test user records; we also filtered some irrelevant actions such as checking the calendar and the instructor profile. From the original 42 actions that Moodle stored by default, we selected only the 16 actions that were relevant to the SRL process and academic performance for the course (Cerezo et al., 2016, 2017).

At this point in preprocessing, as shown by Fayyad and cols. (1996), it is necessary to apply high-level coding to extract useful knowledge from volumes of data and produce meaningful models. That coding must be in accordance with the process discovery goal so in this case according to the assumptions of SRL theory. Based on Zimmerman´s model of SRL, one of the universally accepted, empirically supported models (Zimmerman & Schunk, 2011), we can differentiate three phases during self-regulation of learning: planning, executing and assessment. In addition to this we wanted to monitor forum-supported collaborative learning since forum behavior has been previously shown to be positively related to learning in LMSs (Romero, López, Luna & Ventura, 2013), as has the importance of co-regulated, and socially shared regulation of learning (Hadwin, Järvelä, & Miller, 2011). The resulting high-level coding has five action labels (Planning, Learning, Executing, Review and Forum Peer Learning) (See Table 1). Based on this, we will consider the student as the case and the union between action and high level coding attributes as the event classes. Therefore, each row in the preprocessed event logs is an event class (action and high level coding attributes), carried out by a case (student) on a specific date (timestamp).

**Table 1.** High level coding of the attribute *actions.*

| **Low level Moodle Action** | **High Level Coding** |
|---|---|
| assign submit | EXECUTING |
| assign view | PLANNING |
| forum add discussion | FORUM PEER LEARNING |
| forum add post | FORUM PEER LEARNING |
| forum update post | FORUM PEER LEARNING |
| forum view discussion | FORUM PEER LEARNING |
| forum view forum | FORUM PEER LEARNING |
| page view | LEARNING |
| quiz attempt | EXECUTING |
| quiz close attempt | EXECUTING |
| quiz continue attempt | EXECUTING |
| quiz review | REVIEW |
| quiz view | PLANNING |
| quiz view summary | PLANNING |
| resource view | LEARNING |
| URL view | LEARNING |

Subsequently, the log file was split into two groups based on students' final marks: Pass (containing only events of students who passed the course) and Fail (containing only events of students who failed the course). To do that, we transformed each student's final mark (a numerical value on a 10-point scale) into a categorical value using traditional Spanish academic grading: Fail from 0 to 4.9 and Pass from 5 to 10. We checked that clustering by marks during preprocessing was useful for improving both the performance and comprehensibility of the PM models (Bogarín, Romero, Cerezo, & Sánchez-Santillán, 2014; Romero & Ventura, 2013).

Apart from the Pass-Fail files, we also increased the granularity and divided the event files into sub-files by unit in order to analyze student behavior more thoroughly. In our case, the course was made up of 11 units with different content but the same instructional design. For this reason, we used the information attribute in each record in order to ascertain which unit it belonged to. Table 2 shows the final number of cases and number of events in each unit after preprocessing.

**Table 2.** Number of cases and events per files and units at the datasets.

| Files | Number of Cases | Number of Events |
|---|---|---|
| Group Pass | 73 | 15637 |
| Group Fail | 28 | 5992 |
| Unit 1 | 101 | 1782 |
| Unit 2 | 101 | 2103 |
| Unit 3 | 100 | 2192 |
| Unit 4 | 101 | 2946 |
| Unit 5 | 100 | 2514 |
| Unit 6 | 101 | 1612 |
| Unit 7 | 95 | 2067 |
| Unit 8 | 87 | 1931 |
| Unit 9 | 86 | 1699 |
| Unit 10 | 87 | 1163 |
| Unit 11 | 84 | 1620 |

*Procedure*

The study took place during a one semester assignment which was part of a compulsory 3rd year subject completed entirely outside teaching hours. Students were informed about the data collection and gave their informed consent. The course was made up of 11 lessons that were delivered to the students on a weekly basis. They were asked participated in eTraining about study strategies related to the subject topic (Cerezo, Núñez, Rosario, Valle, Rodríguez, & Bernardo, 2010; Núñez et al., 2011). Following Biggs (2005), each lesson was composed of three different types of content:

- Declarative knowledge level, theoretical content description, information, and knowledge of how to put into practice the strategies of the week.
- Procedural knowledge level, practical tasks where the students put the declarative knowledge into practice.
- Conditional knowledge level, discussion forums where the students discuss how they have used or would use the strategy or strategies of the week in different contexts.

The instructor strongly suggested that students approached the assignments for each unit in the following order: understand the theoretical content, put them into practice through the corresponding task, and share their experience about the week’s topic in the forum; a learning path supported by the instructional design and SRL theory. The content of each course (whether theoretical, practical or forum related) was designed through Moodle resources to ensure that the student had to interact with the system during their learning experience, e.g. avoiding downloading files. In this way we aimed to ensure that the learners were leaving the traces that are used in this study in the form of logs. In any case, the students were free to follow their own learning path and the only compulsory assignments for each unit were to complete the weekly practical task and to post at least one comment in each unit forum. It was estimated that each unit required an average of 2.5-3 hours of student work per week, including comprehensive reading, practical tasks and forums.

*Data analysis*

Data analysis involved the log file preprocessing and transformation into the XES (eXtensible Event Stream) file which is required to implement process mining using the well-known ProM framework, then the process discovery, and finally the interpretation of the model (Romero, Cerezo, Bogarín, & Sánchez-Santillán, 2016) See Figure 1.

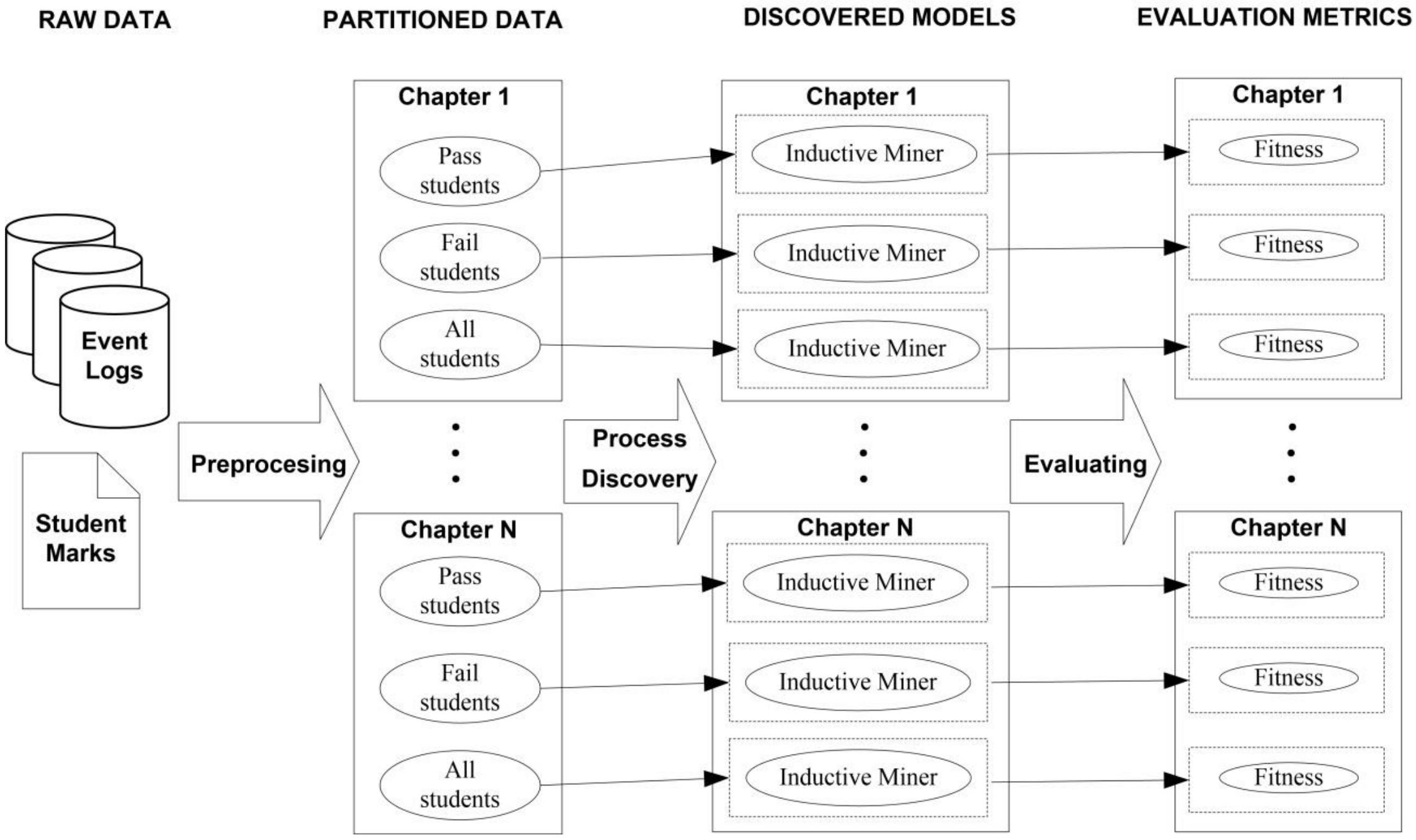

**Figure 1.** EPM process from raw data to algorithm interpretation.

In order to model our datasets, we executed the selected PM algorithm Inductive Miner for each file. To test how well the obtained models described the observed data we used the fitness evaluation metric, which quantifies the extent to which the discovered model can accurately reproduce the cases recorded in the log. There are other alternative quality indexes important for process discovery such as precision and generalization, however, it only makes sense to consider the other indexes if fitness is acceptable (Buijs et al., 2012; Van der Aalst, 2016), otherwise what we obtain in the model would not represent reality. The greatest importance has also been placed on fitness in previous studies (Bogarin et al, 2018; Buijs et al., 2012).

RESULTS

Table 3 shows the results of the IM algorithms in the *fitness* evaluation metric. In general IM algorithm fitness scored higher by units than when looking at the whole course (all units together). The same table also shows the improving effect of grouping the data by dividing logs into Pass and Fail students instead of using all the students.

**Table 3.** Fitness for the IM by grouping data.

| Units | Pass students | Fail students | All students |
|---|---|---|---|
| All Units | 0.872 | 0.881 | 0.659 |
| Unit 1 | 0.947 | 0.935 | 0.797 |
| Unit 2 | 0.966 | 0.921 | 0.781 |
| Unit 3 | 0.942 | 0.975 | 0.712 |
| Unit 4 | 0.959 | 0.870 | 0.747 |
| Unit 5 | 0.889 | 0.943 | 0.749 |
| Unit 6 | 0.861 | 0.926 | 0.793 |
| Unit 7 | 0.893 | 0.967 | 0.773 |
| Unit 8 | 0.877 | 0.906 | 0.796 |
| Unit 9 | 0.911 | 0.868 | 0.784 |
| Unit 10 | 0.975 | 0.978 | 0.856 |
| Unit 11 | 0.987 | 0.938 | 0.778 |

Along with quality metrics, Figures 2 and 3 show the two obtained models for sub-files Pass and Fail when using All Units of the course together. In order to understand and interpret these IM-generated models it is necessary to understand what each visual element means. The boxes are the activities carried out by the students, the number in the box is the frequency, the arrows indicate the direction of the process, the number above the arrows is the frequency of the transition between these two actions, and the diamonds with a cross represent parallelism. Each model begins with an initial node and ends with a final node.

In addition to exhibiting lower fitness than by units, the models for the Pass and Fail subfiles show huge parallelism, which means that they are not clearly detailing the route followed by the students. In Figure 2, apart from two isolated forum actions, the rest of the Fail students' behavior is not represented as a process but as a collection of actions. The same happens in case of the Pass students (Figure 3), apart from some initial *forum peer learning* we find a very parallel model where interpretation is not possible.

The frequency and relationships of events is greater in the graph of all the Pass and all the Fail students than in the ones obtained when data are grouped by units. Better measurements were obtained in a dataset with fewer records as seen in Table 3 because the number of records in the file is lower (back to Table 2). As well as this, the model obtained for Pass group does not represent the workflow that all the passing students performed on the platform, showing only 67 out of 73 students´ behaviour on the platform. Although the Unit instructional design and instructor's suggestions were the same, the visualization erred on the side of simplification.

For the aforementioned reasons, we planned to increase the granularity and produced sub-files by unit in order to analyse student behaviour more thoroughly.

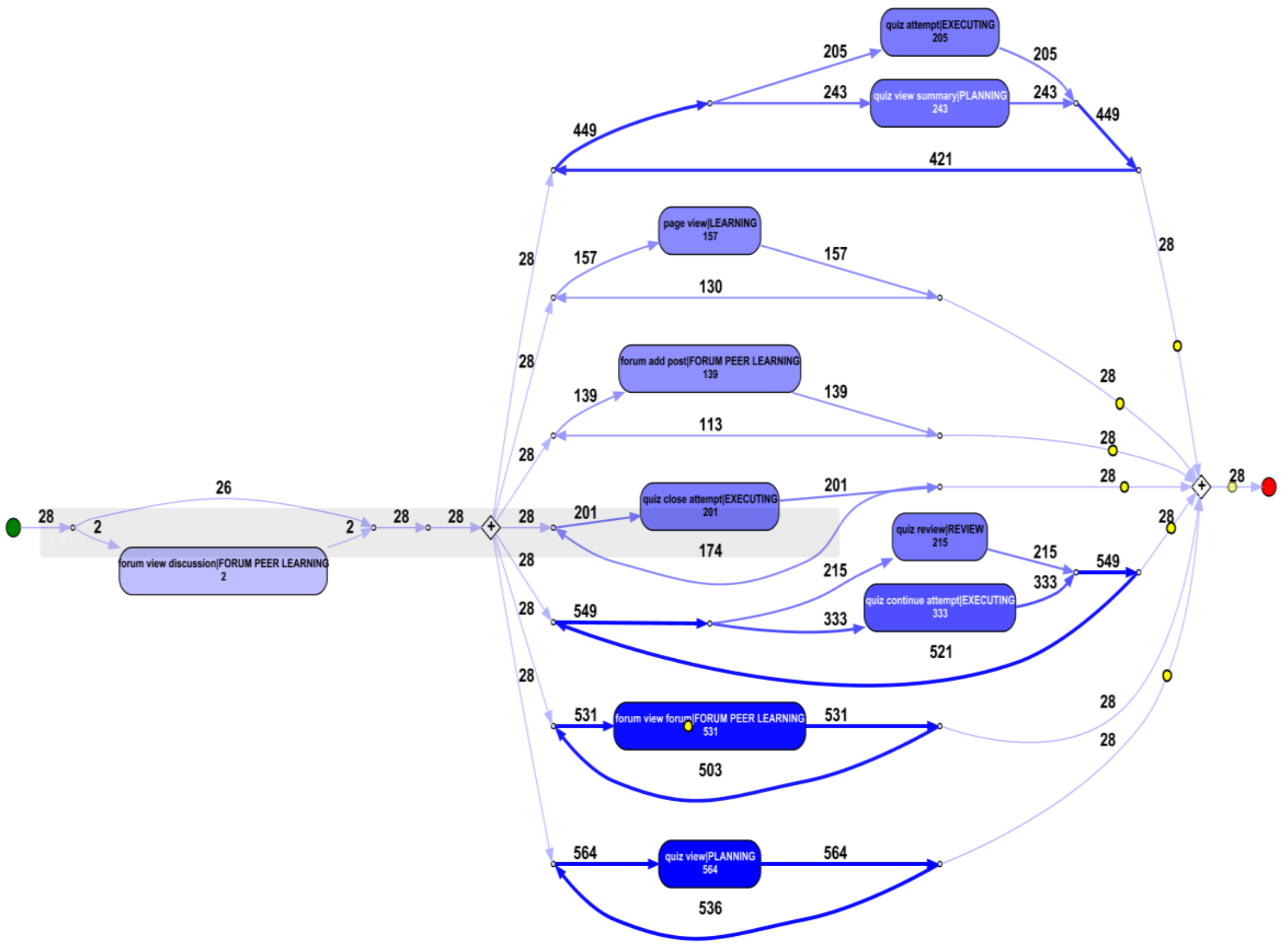

**Figure 2.** Visualization of failing students' learning path in All Units sub-file Pass

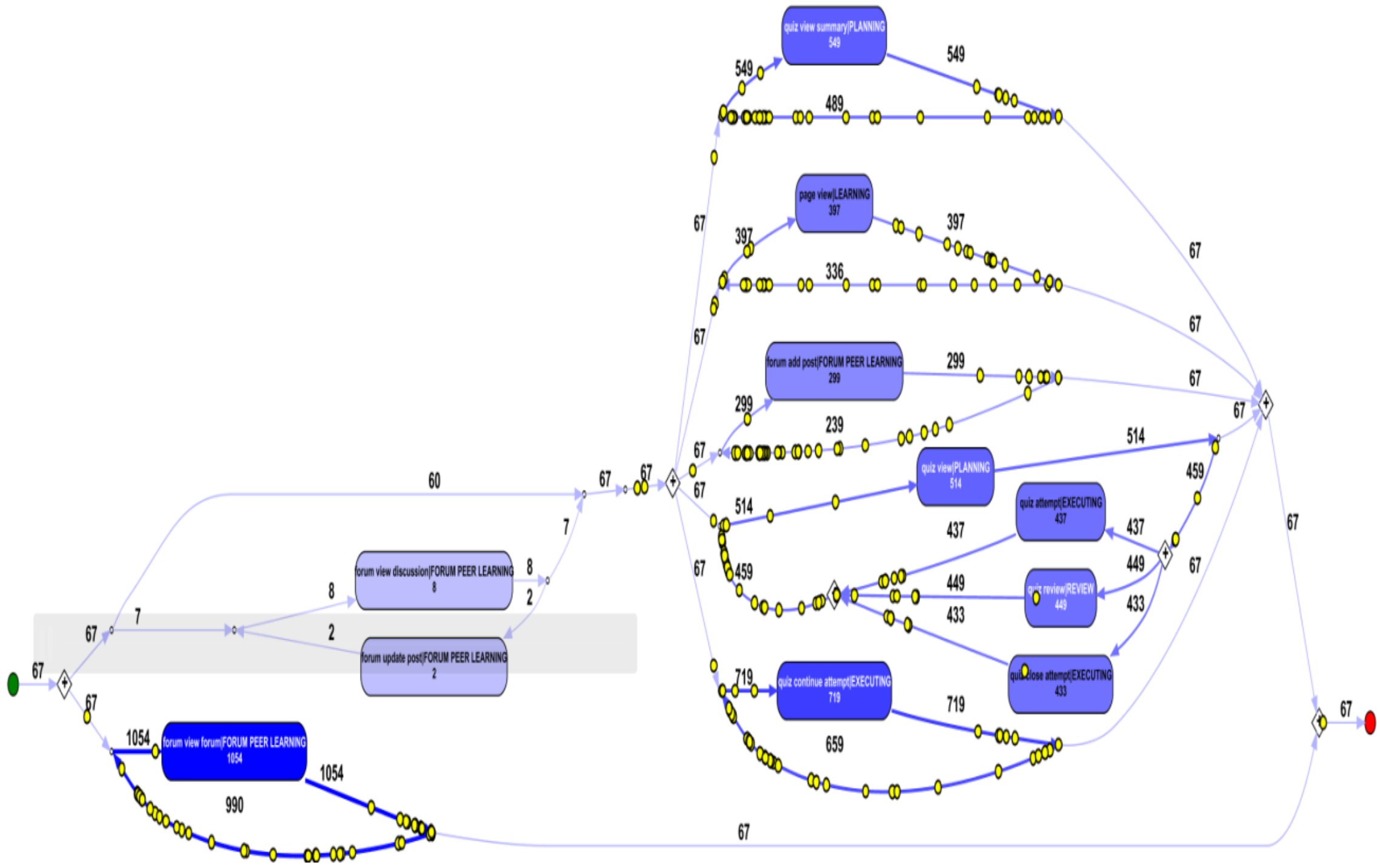

**Figure 3.** Visualization of failing students' learning path in All Units sub-file Fail

Figures 4 and 5 (the figures have been split to improve their visualization) show two examples of obtained models when using only one unit (in this case sub-file or unit 4). This unit was chosen because is where the students had more interaction with the LMS resulting in a higher number of cases and events (back to Table 2).

The networks by unit are longer and more developed than by all units together. They show a more structured student workflow in contrast to the networks previously obtained with the complete course that were more flattened.

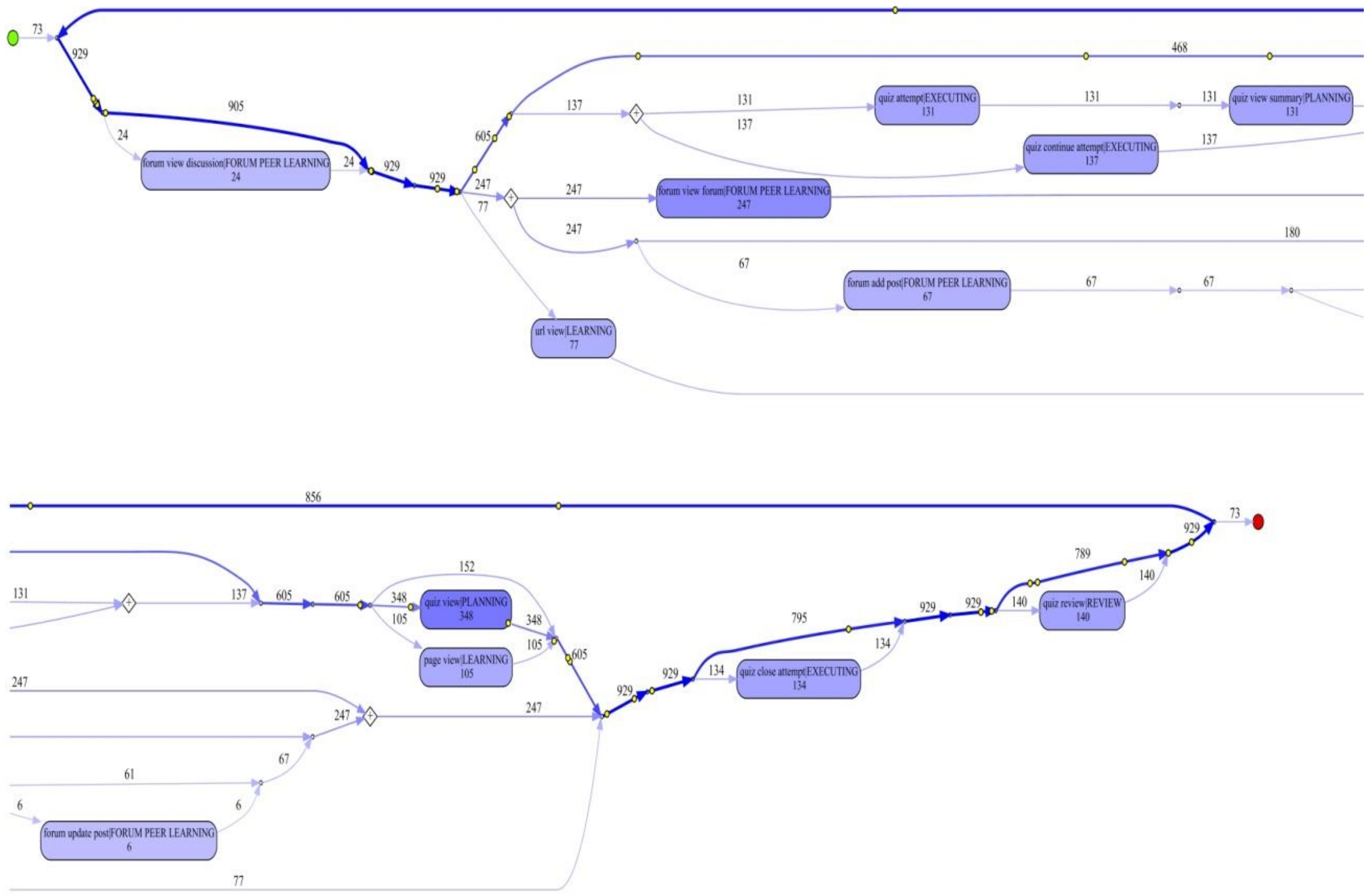

**Figure 4.** Visualization of passing students' learning path in Unit sub-file 4

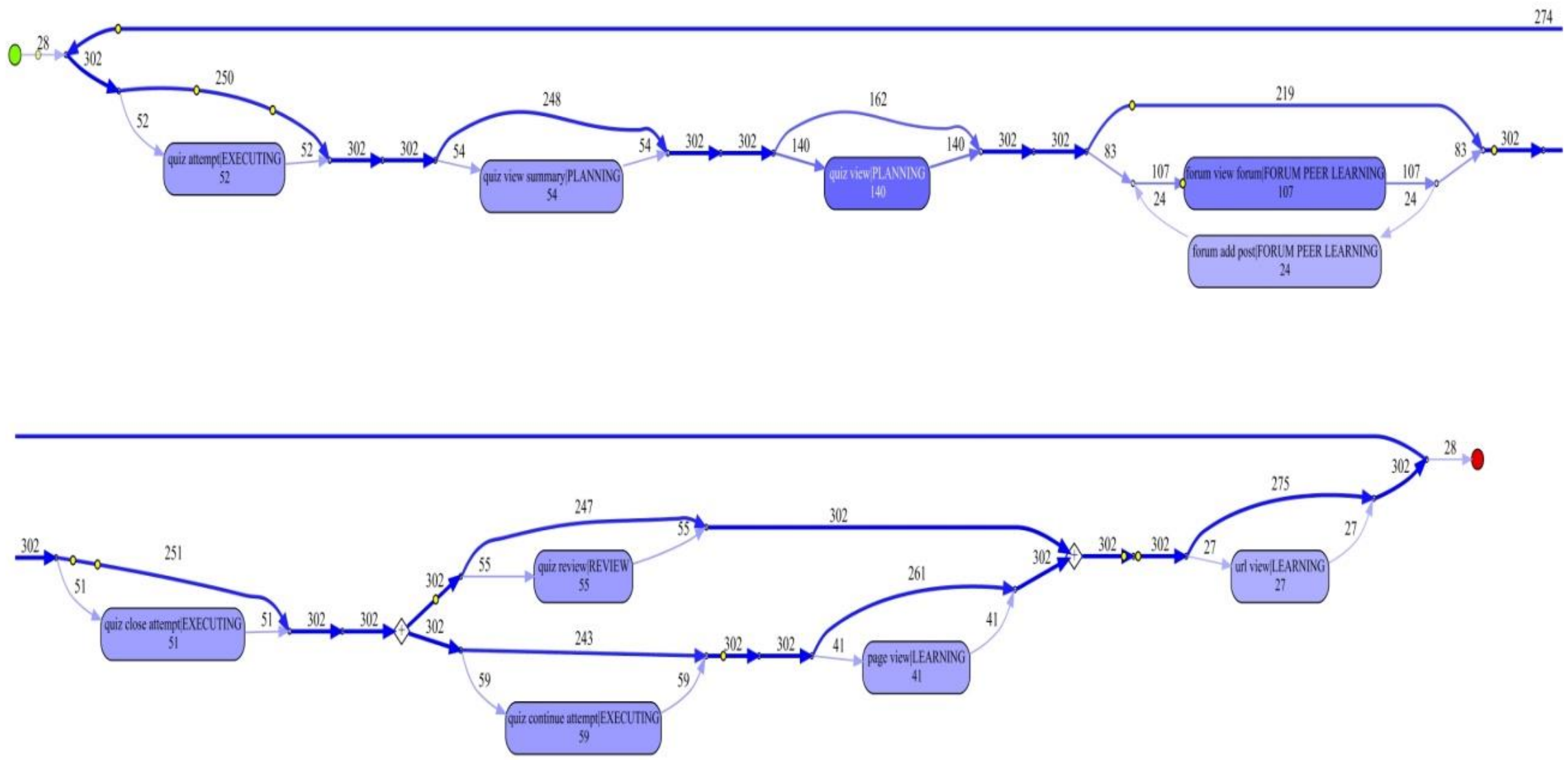

**Figure 5.** Visualization of failing students' learning path in Unit sub-file 4

In this case, students in the Pass group (Figure 4) started their study process by visiting the *forum view discussion,* after which the model splits into different possible routes. One route continues via the *URL view*, the second route involves continuing the study process with *forum view forum, forum add post* or *forum update post.* There is also a third route, in which student perform actions related to the quizzes *quiz attempt, quiz view summary* and *quiz continue attempt.* The extracted model ends with the *quiz close attempt* and *quiz review* actions. Students in the Fail group (see Figure 5), show *quiz attempt at* the beginning, followed by the *quiz view summary* and the *quiz view*. All of the activities are related to the compulsory course assignments. In the middle part of the model, they do forum-related activities such as *forum view forum* and *forum add post*, which is the other compulsory activity. Following that there are parallel actives *-quiz review, quiz continue attempt-,* finishing with *page view* and *URL view* which would have been the logical starting point suggested by the instructor.

CONCLUSIONS

The aim of this study was to assess students' SRL skills during an e-Learning course through EPM techniques. To do that, we analyzed a log file with 21,629 events from an on-line course for Spanish undergraduate students. The study involved preprocessing to implement process mining, followed by process discovery and algorithm interpretation. In order to discover models from our data, we used the algorithm Inductive Miner as a novel technique applied to educational data. To test how well the obtained model described the observed data, we extracted the fitness to give priority to both the extent to which the discovered model could accurately reproduce the cases recorded in the log, and practical utility for researchers and instructors.

Based on the results, we can draw two important conclusions. Firstly, the IM algorithm produces models with good fitness values which means that it correctly reproduces the students' interactions on the Moodle platform. We also achieved better results when we divided the data by units. This make sense since better measurements are obtained in a dataset with fewer records as observed in previous studies (Bogarín et. al., 2018; Bogarin et al., 2014.). In any case, it seems that applying the IM algorithm to discovering SRL behavior models opens a new field in the assessment and understanding of skills in e-learning. It is essential since face-to-face learning skill evaluation in general, and SRL in particular, is itself a challenge in educational sciences. In this sense, the electronic learning environment along with Process Mining techniques are positioned as promising solutions in this research field.

Secondly, giving priority to practical utility in authentic contexts is essential to achieve the desired connection between research and practice (Cerezo, R., Fernández, E., Amieiro, N., Valle, A., Rosário, P., & Núñez, J. C. (2018). EDM in general and EPM in particular are tools that require high technical knowledge. Along with this, PM algorithms often result in uninterpretable or spaghetti-like process models (Van der Aalst, 2011) which are very hard to read with or without technical knowledge. To that end, apart from applying a high-level coding scheme based on self-regulated students' behavior for improving the comprehensibility of EPM models (Cerezo, Romero, Bogarín, & Núñez, 2014; Zimmerman, 1990), we wanted to obtain meaningful models that could be interpreted. Increasing the granularity and producing sub-files by unit made it possible to analyze student behavior more thoroughly and produced sounder models which were easier to interpret. The networks by unit sub-file were longer and more developed, showing a more structured student workflow so instructors would be able to visualize and interpret the behavior of students' learning paths.

In this regard, if we look at the high level coding of the failing cluster in Unit 4, the model shows that the students who failed did not follow the learning path suggested by the instructor and supported by self-regulatory skills. However, if we look at the high-level coding of students who passed, we can see that they did not follow the instructors' suggestions exactly;

Lust, Elen, and Clarebout (2013) previously observed that only a minority of students regulated their behavior in line with course requirements. Nevertheless, they exhibited a much more meaningful and self-regulated learning process. Passing students started with actions indicating comprehension and learning of the materials and concluded with executing and reviewing actions. Additionally, the model let us detect that the Pass cluster performed actions related to forum-supported collaborative learning which do not appear in the model from the Fail cluster. An interesting finding in line with previous studies that found relationships between forum behavior and student achievement in LMSs (Romero, López, Luna & Ventura, 2013).

Going further, the visualization of learning analytics is essential. Visual analytics of different learning models could be helpful in making appropriate, real time decisions during the teaching-learning process (Duval, 2011; Gómez-Aguilar, Hernández-García, García-Penñalvo & Theron, 2015).

A feasible application of the results as a whole also concerns early prediction of failure and difficulties, a very promising area of study (Hu, Lo, & Shih, 2014; Wolff, Zdrahal, Herrmannova, Kuzilek, & Hlosta, 2014). Modeling behavior with IM algorithms could contribute to developing early warning systems to predict at-risk students while a course is in progress. As well as this, these results could contribute to the personalization of e-learning environments in terms of self-regulation building Recommendation Systems based on different SRL behaviors having to be stressed in one of the SRL phases or evaluating the adequacy of different types of prompts for different SRL behavior models (Lehmann, Hähnlein, & Ifenthaler, 2014).

Finally, although the results of this study appear robust, the data were provided by third-year graduate students, so it is possible that the models would be different in the case of first-year graduate students. It was found that novice students reported less sophisticated study SRL strategies to address new domains of information (Ge & Harde, 2010). As well as this, it is possible that the results may vary based on students' degrees although the tasks themselves were unrelated to the degree.

In addition, LMS are considered just another component of the learning ecosystem (García-Peñalvo & Seoane, 2015). This raises awareness about the future work on changing the focus to other relevant virtual learning environments, such as Personal Learning Environments (PLEs) or Massive Open Online Courses (MOOCs), and verifying our findings across different types of learning platforms.

In summary, this study, albeit with those limitations, aimed to shed some light on the e-teaching-e-learning process through EPM techniques and to be useful to the fundamental participants in the teaching-learning process, teachers and learners.

## ACKNOWLEDGMENTS

This work was funded by the Department of Science and Innovation (Spain) under the National Program for Research, Development and Innovation: project TIN2017-83445-P. We have also received funds from the European Union, through the European Regional Development Funds (ERDF); and the Principality of Asturias, through its Science, Technology and Innovation Plan FC-GRUPIN-IDI/2018/000199.

## REFERENCES

Aljawarneh, S., Muhsin, Z., Nsour, A., Alkhateeb, F., & AlMaghayreh, E. (2010, May). E-learning tools and technologies in education: a perspective. In *The Fifth International Conference of Learning International Networks Consortium (LINC), MIT. Cambridge, MA. Retrieved from: http://people. math. sfu. ca/~ vjungic/shadi. pdf.*

Azevedo, R., & Aleven, V. (Eds.). (2013). *International handbook of metacognition and learning technologies.* Amsterdam, The Netherlands: Springer.

Azevedo, R., & Feyzi-Behnagh, R. (2011). Dysregulated learning with advanced learning technologies. *Journal of e-Learning and Knowledge Society*, *7*(2), 9e18.

Azevedo, R., Harley, J., Trevors, G., Duffy, M., Feyzi-Behnagh, R., Bouchet, F., & Landis, R. (2013). Using trace data to examine the complex roles of cognitive, metacognitive, and emotional self-regulatory processes during learning with multi-agent systems. In R. Azevedo, & V. Aleven(Ed.), *International handbook of metacognition and learning technologies* (pp. 427-449). Springer New York.

Bannert, M., Reimann, P., & Sonnenberg, C. (2014). Process mining techniques for analysing patterns and strategies in students' self- regulated learning. In: *Metacognition and learning*, *9*(2), 161-185. doi:10.1007/s11409-013-9107-6

Bernard, R. M., Abrami, P. C., Lou, Y., Borokhovski, E., Wade, A., Wozney, L., ... & Huang, B. (2004). How does distance education compare with classroom instruction? A meta-analysis of the empirical literature. *Review of educational research*, *74*(3), 379-439.

Biggs, J. B. (2005). *Calidad del aprendizaje universitario* [Quality of university learning]. Madrid: Narcea.

Bjork, R. A., Dunlosky, J., & Kornell, N. (2013). Self-regulated learning: Beliefs, techniques, and illusions. *Annual review of Psychology, 64*, 417-444. https://doi.org/10.1146/annurev-psych-113011-143823

Bogarín, A., Cerezo, R., & Romero, C. (2018). A survey on educational process mining. *Wiley Interdisciplinary Reviews: Data Mining and Knowledge Discovery*, *8*(1). doi:10.1002/widm.1230

Bogarín, A., Romero, C., Cerezo, R., & Sánchez-Santillán, M. (2014). Clustering for improving educational process mining. In M. Pistilli, J. Willis, & D. Koch (Eds.), *Proceedings of the Fourth International Conference on Learning Analytics And Knowledge* (pp. 170-181). Indianapolis, USA: ACM. doi:10.1145/2567574.2567604

Buijs, J. C., Van Dongen, B. F., & van Der Aalst, W. M. (2012). On the role of fitness, precision, generalization and simplicity in process discovery. In R. Meersman, H. Panetto, T. Dillon, S. Rinderle-Ma, P, Dadam, X. Zhou, S. Pearson, A. Ferscha, S. Bergamaschi, & I. F. Cruz, *Proceedings of the OTM Confederated International Conferences" On the Move to Meaningful Internet Systems"* (pp. 305-322). Berlin: Springer. doi:10.1007/978-3-642-33606-5_19

Cerezo, R., Esteban, M., Sánchez-Santillán, M., & Núñez, J. C. (2017). Procrastinating Behavior in Computer-Based Learning Environments to Predict Performance: A Case Study in Moodle. *Frontiers in psychology*, *8*, 1403.

Cerezo, R., Romero, C., Bogarín, A., & Núñez J.C. (2014) Improving performance and comprehensibility of Educational Process Mining models for a better understanding of the learning process. Metacognition 2014. 6th Bienal Meeting of the EARLI Special Interest Group 16. Estambul, Turquia, (pp. 1-2).

Cerezo, R., Sánchez-Santillán, M., Paule-Ruiz, M. P., & Núñez, J. C. (2016). Students' LMS interaction patterns and their relationship with achievement: A case study in higher education. *Computers & Education*, *96*, 42-54. doi:10.1016/j.compedu.2016.02.006

Cerezo, R., Fernández, E., Amieiro, N., Valle, A., Rosário, P., & Núñez, J. C. (2018). Mediating Role of Self-efficacy and Usefulness Between Self-regulated Learning Strategy Knowledge and its Use. *Revista de Psicodidáctica*. doi: 10.1016/j.psicod.2018.08.001

Cerezo, R., Nuñez, J. C., Rosario, P., Valle, A., Rodriguez, S., & Bernardo, A. (2010). New Media for the promotion of self-regulated learning in higher education. Psicothema, 22(2), 306-315.

Clark, R. C., & Mayer, R. E. (2016). *E-learning and the science of instruction: Proven guidelines for consumers and designers of multimedia learning*. New Jersey: John Wiley & Sons.

Cook, D. A., Levinson, A. J., Garside, S., Dupras, D. M., Erwin, P. J., & Montori, V. M. (2008). Internet-based learning in the health professions: a meta-analysis. *Jama*, *300*(10), 1181-1196.

Dahlstrom, E., Brooks, D. C., & Bichsel, J. (2014). *The current ecosystem of learning management systems in higher education: Student, faculty, and IT perspectives* (Research report). Retrieved from http:// www. educause. edu/ecar. 2014 EDUCAUSE. CC by-nc-nd

Duval, E. (2011, February). Attention please!: learning analytics for visualization and recommendation. In *Proceedings of the 1st international conference on learning analytics and knowledge* (pp. 9-17). ACM. doi: 10.1145/2090116.2090118

Emond, B., & Buffett, S. (2015, June). *Analyzing Student Inquiry Data Using Process Discovery and Sequence Classification.* Paper presented at the International Educational Data Mining Society, Madrid, Spain.

European Commission. (2014). *New modes of learning and teaching in higher education.* Luxembourg: European Union.

Fayyad, U., Piatetsky-Shapiro, G., & Smyth, P. (1996). The KDD process for extracting useful knowledge from volumes of data. *Communications of the ACM*, *39*(11), 27-34. doi: 10.1145/240455.240464

García-Peñalvo, F. J., & Seoane Pardo, A. M. (2015). Una revisión actualizada del concepto de eLearning. *Educ. Knowl. Soc. 16*, 119–144. doi: 10.14201/eks2015161119144

Ge, X., & Harde, P. L. (2010). Self-processes and learning environment as influences in the development of expertise in instructional design. Learning Environment Research, 13(1), 23e41.

Gómez-Aguilar, D. A., Hernández-García, A., García-Penñalvo, F. J., & Theron, R. (2015). Tap into visual analysis of customization of grouping of activities in eLearning. *Computers in Human Behavior, 47*, 60e67.

Hadwin, A. F., Järvelä, S., & Miller, M. (2011). Self-regulated, co-regulated, and socially shared regulation of learning. *Handbook of self-regulation of learning and performance*, *30*, 65-84.

Hattie, J. (2008). *Visible learning: A synthesis of over 800 meta-analyses relating to achievement*. Oxon, OX: Routledge.

Hu, Y. H., Lo, C. L., & Shih, S. P. (2014). Developing early warning systems to predict students' online learning performance. *Computers in Human Behavior*, *36,* 469-478. https://doi.org/10.1016/j.chb.2014.04.002

Lara, J. A., Lizcano, D., Martínez, M. A., Pazos, J., & Riera, T. (2014). A system for knowledge discovery in e-learning environments within the European Higher Education Area–Application to student data from Open University of Madrid, UDIMA. *Computers & Education*, *72*, 23-36.

Lehmann, T., Hähnlein, I., & Ifenthaler, D. (2014). Cognitive, metacognitive and motivational perspectives on preflection in self-regulated online learning. *Computers in Human Behavior, 32,* 313e323. doi:10.1016/j.chb.2013.07.051.

Lust, G., Elen, J., & Clarebout, G. (2013a). Regulation of tool-use within a blended course: student differences and performance effects. *Computers & Education*, *60*(1), 385-395.

Means, B., Toyama, Y., Murphy, R., Bakia, M., & Jones, K. (2009). Evaluation of evidence-based practices in online learning: A meta-analysis and review of online learning studies. Jessup, MD: US Department of Education

Merchant, Z., Goetz, E. T., Cifuentes, L., Keeney-Kennicutt, W., & Davis, T. J. (2014). Effectiveness of virtual reality-based instruction on students' learning outcomes in K-12 and higher education: A meta-analysis. *Computers & Education, 70,* 29-40. doi:10.1016/j.compedu.2013.07.033

Núñez, J. C., Cerezo, R., Bernardo, A., Rosário, P., Valle, A., Fernández, E., et al. (2011). Implementation of training programs in self-regulated learning strategies in moodle format: results of a experience in higher education. Psicothema, 23(2), 274e281.

Paechter, M., Maier, B., & Macher, D. (2010). Evaluation universitärer Lehre mittels Einschätzungen des subjektiven Kompetenzerwerbs. *Psychologie in Erziehung und Unterricht*, *2,* 128-138.

Reimann, P., Markauskaite, L., & Bannert, M. (2014). E-Research and learning theory: What do sequence and process mining methods contribute? *British Journal of Educational Technology*, *45*(3), 528- 540. doi:10.1111/bjet.12146

Romero, C., & Ventura, S. (2013). Data mining in education. *Wiley Interdisciplinary Reviews: Data Mining and Knowledge Discovery, 3*(1), 12-27.

Romero, C., López, M. I., Luna, J. M., & Ventura, S. (2013). Predicting students' final performance from participation in on-line discussion forums. *Computers & Education, 68,* 458-472.

Sanchez-Santillan, M., Paule-Ruiz, M., Cerezo, R., & Alvarez-García, V. (2016). *MeL: Modelo de adaptación dinámica del proceso de aprendizaje en eLearning. anales de psicología, 32*(1), 106-114.

Schmid, R. F., Bernard, R. M., Borokhovski, E., Tamim, R., Abrami, P. C., Wade, C. A., ... & Lowerison, G. (2009). Technology's effect on achievement in higher education: a Stage I meta-analysis of classroom applications. *Journal of computing in higher education*, *21*(2), 95-109.

van der Aalst, W. M. (2011). Process Discovery: An Introduction. In *Process Mining* (pp. 125-156). Berlin, Heidelberg: Springer. doi: 10.1007/978-3-642-19345-3_5

van der Aalst, W. M. (2016). *Process mining: data science in action*. Berlin, Heidelberg: Springer. doi:10.1007/978-3-662-49851-4

Wolff, A., Zdrahal, Z., Herrmannova, D., Kuzilek, J., & Hlosta, M. (2014, March). Developing predictive models for early detection of at-risk students on distance learning modules. *Machine Learning and Learning Analytics Workshop* at The 4th International Conference on Learning Analytics and Knowledge *(LAK14)*, 24-28 Mar 2014, Indianapolis, Indiana, USA. Retrieved from http://lak14indy.wordpress.com/

Zimmerman, B. J. (1990). Self-regulated learning and academic achievement: An overview. *Educational psychologist*, *25*(1), 3-17.

Zimmerman, B. J. (2013). Theories of self-regulated learning and academic achievement: An overview and analysis. In B. J. Zimmerman & D. H. Schunk (Eds.), *Self-regulated learning and academic achievement* (pp. 10-45). London: Routledge.

Zimmerman, B. J., & y Schunk, D. (2011). *Handbook of self-regulation of learning and performance.* New York: Routledge.